\documentclass[prl, twocolumn, preprintnumbers, superscriptaddress, amsmath, amssymb]{revtex4-2}

\usepackage{graphicx}
\usepackage{dcolumn}
\usepackage{bm}
\usepackage{amssymb}
\usepackage{amsmath} 
\usepackage{wasysym}
\usepackage{sidecap}
\usepackage{natbib} 

\begin{document}

\title{Ultrafast optically induced ferromagnetic state in an elemental antiferromagnet}

\author{E. Golias}
\email{evangelos.golias@gmail.com}
\affiliation{Institut f\"ur Experimentalphysik, Freie Universit{\"a}t Berlin, Arnimallee 14, 14195 Berlin, Germany}

\author{I. Kumberg}
\affiliation{Institut f\"ur Experimentalphysik, Freie Universit{\"a}t Berlin, Arnimallee 14, 14195 Berlin, Germany}

\author{I. Gelen}
\affiliation{Institut f\"ur Experimentalphysik, Freie Universit{\"a}t Berlin, Arnimallee 14, 14195 Berlin, Germany}

\author{S. Thakur}
\affiliation{Institut f\"ur Experimentalphysik, Freie Universit{\"a}t Berlin, Arnimallee 14, 14195 Berlin, Germany}

\author{J. G{\"o}rdes}
\affiliation{Institut f\"ur Experimentalphysik, Freie Universit{\"a}t Berlin, Arnimallee 14, 14195 Berlin, Germany}

\author{R. Hosseinifar}
\affiliation{Institut f\"ur Experimentalphysik, Freie Universit{\"a}t Berlin, Arnimallee 14, 14195 Berlin, Germany}

\author{Q. Guillet}
\affiliation{Institut f\"ur Experimentalphysik, Freie Universit{\"a}t Berlin, Arnimallee 14, 14195 Berlin, Germany}

\author{J. K. Dewhurst}
\affiliation{Max-Planck-Institut f\"ur Mikrostrukturphysik, Weinberg 2, 06120 Halle, Germany}

\author{S. Sharma}
\affiliation{Max Born Institute for Nonlinear Optics and Short Pulse Spectroscopy, Max-Born-Strasse 2A, 12489 Berlin, Germany}

\author{C. Sch{\"u}{\ss}ler-Langeheine}
\affiliation{Helmholtz-Zentrum Berlin f\"ur Materialien und Energie, Albert-Einstein Stra{\ss}e 15, 12489 Berlin, Germany}

\author{N. Pontius}
\affiliation{Helmholtz-Zentrum Berlin f\"ur Materialien und Energie, Albert-Einstein Stra{\ss}e 15, 12489 Berlin, Germany}

\author{W. Kuch}
\affiliation{Institut f\"ur Experimentalphysik, Freie Universit{\"a}t Berlin, Arnimallee 14, 14195 Berlin, Germany}

\begin{abstract}

We present evidence for an ultrafast optically induced ferromagnetic alignment of antiferromagnetic Mn in Co/Mn multilayers. We observe the transient ferromagnetic signal at the arrival of the pump pulse at the Mn L$_3$ resonance using x-ray magnetic circular dichroism in reflectivity. The timescale of the effect is comparable to the duration of the excitation and occurs before the magnetization in Co is quenched. Theoretical calculations point to the imbalanced population of Mn unoccupied states caused by the Co interface for the emergence of this transient ferromagnetic state.

\end{abstract}

\maketitle


Controlling magnetic order at high speeds requires the ultrafast manipulation of the spin degree of freedom, a central goal of spintronics \cite{Zutic:2004sp}. Progress in lasers rendered ultrashort optical pulses as the most promising route towards the ultrafast control of magnetization \cite{Kirilyuk:2010rm}. The importance for technological applications and the scientific interest for the physical processes underlying ultrafast demagnetization focused a lot of research on ultrafast optical quenching of magnetic order in itinerant ferromagnetic (FM) materials after a non-adiabatic excitation at timescales comparable or even shorter than the exchange interaction \cite{Beaurepaire:1996vo,Koopmans:2009ds,Boeglin:2010dz,Rudolf:2012ny,Eschenlohr:2013id,Turgut:2013kx,Turgut:2016dx,Eich:2017im,Shokeen:2017jd,Gort:2018eb,Maldonado:2020ku}.

On the contrary, reports on itinerant antiferromagnets are scarce because the absence of a macroscopic magnetic moment makes these systems difficult to study. Lately, \citet{Thielemann:2017nh} showed that manipulation of antiferromagnetic (AFM) order is considerably faster than FM, a pivotal finding given the modern perspective for the cooperative utilization of FM and AFM components in future (opto-) spintronics \cite{Baltz:2018fw, Nemec:2018af}. In this context, materials that can switch between AFM and FM order on ultrafast timescales could offer unprecedented opportunities. Such a transition has been observed in the time domain for the first time in FeRh \cite{Ju:2004gs, Thiele:2004gr} on sub-ps timescale after excitation with fs laser pulses. Later, Radu et al. \cite{Radu:2011kr} reported on the formation of a transient FM state at the ps timescale during the magnetization reversal of the ferrimagnetic material GdFeCo.

A critical question is: how fast can we induce such a transient FM state in an AFM? In FeRh, the laser pulse heats the electronic system with the concomitant modification of the exchange field that couples the spins antiferromagnetically. Afterwards, the FM state emerges because of a Rh-mediated strong FM exchange interaction of Fe atoms \cite{Ju:2004gs}. In the case of GdFeCo, the competition between thermal and exchange energy can transiently drive the two sublattices to a FM alignment \cite{Khorsand:2012bt}. Eventually, the lowest temporal limit for thermally activated processes, common in itinerant systems, is set by the timescale of the exchange interaction ($\lessapprox$ 100 fs) \cite{Pajda:2001mr}. Nevertheless, as the electronic response to an electric field is virtually instantaneous, optical excitations might allow for the control of magnetic order at timescales shorter than the exchange interaction \cite{Zhang:2000bl,Gomez-Abal:2004gl}.

A mechanism that enables the all-optical manipulation of magnetic order on sub-exchange timescales is the optically induced intersite spin transfer (OISTR) \cite{Dewhurst:2018jt}. It is of pure optical origin, as spin-selective transfer is taking place between neighboring atoms driven by the oscillating electric field of light. The process is universal, i.e., it does not depend on the material, and allows control of magnetic order only with the structure of the excitation pulse. After its theoretical prediction \cite{Dewhurst:2018jt}, experiments using time-resolved magnetic circular dichroism with extreme ultraviolet photons in Ni/Pt multilayers confirmed the presence of OISTR in the ultrafast demagnetization of the FM Ni layer \cite{Siegrist:2019ii}, opening the way for the magnetic control on attosecond timescales, an order of magnitude faster than the exchange interaction. Shortly afterwards, other experimental studies concluded the existence of OISTR at the Co/Cu(001) interface \cite{Chen:2019fe} using second-harmonic generation and by tracing the demagnetization in CoPt \cite{Willems:2020hd} and FeNi alloys \cite{Hofherr:2020dv} using  extreme ultraviolet magnetic circular dichroism and transverse Kerr effect, respectively. Nevertheless, the most intriguing prediction of OISTR \cite{Dewhurst:2018jt} is yet to be observed: an ultrafast optically-driven transient FM state in an AFM material.

\begin{figure*}
\includegraphics [width=0.75\textwidth]{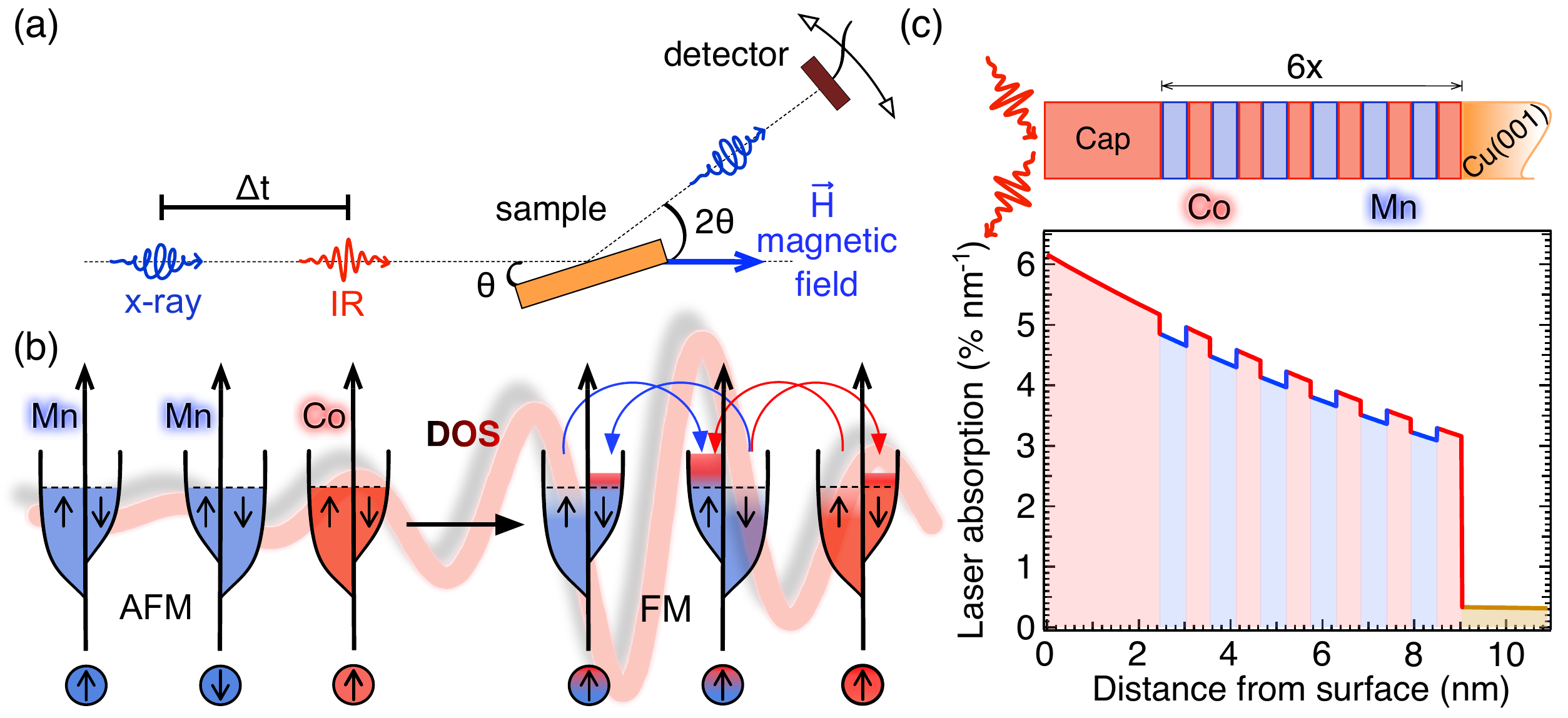}
\caption{(a) Schematic representation of our experimental geometry. The intensity of the reflected x-rays is probed by a photodiode in a $\Theta-2\Theta$ specular geometry with incidence angle $\theta$ = 8.75$^o$. Oscillating red and spiraling blue arrows indicate the linearly-polarized infrared laser and the circularly-polarized soft-x-ray pulses, respectively. (b) The underlying principle of the OISTR mechanism: the electric field of light can, above a threshold value, transfer electrons coherently between atoms. In an AFM material, majority states from the first atom are transferred to the minority of the second (and vice-versa). When an asymmetry is present, such as the interface between Mn and Co, a charge filling imbalance is introduced and a transient FM alignment emerges. (c) Calculation of the differential laser power absorption on each layer of the Co/Mn multilayer. Red and blue lines correspond to the absorption from Co and Mn layers, respectively. The gold line corresponds to the absorption from the Cu(001) substrate, here, only a small part close to the interface with the multilayer is shown. The absorption in the substrate slowly becomes virtually zero within 40 nm from the interface. On the upper part, a schematic of our sample is displayed. Red oscillating arrows represent the incoming and reflected laser light, while red, blue, and gold blocks represent the Co, Mn layers and the Cu(001) substrate, respectively.}
\label{fig1}
\end{figure*}

In this Letter, we report the observation of such a transient FM state in the AFM Mn in a Co/Mn multilayer, after an ultrashort laser excitation. Our sample consists of repetitions of 3 monolayers (MLs) of Mn and 3 MLs of Co, in which, under static conditions, the magnetizations of the FM Co layers are FM aligned and the net Mn magnetization of the AFM Mn layers is close to zero. We unambiguously observe a transient FM state in the AFM Mn by magnetic circular dichroism in time-resolved resonant magnetic x-ray reflection (RMXR), which is estimated to last as long as the pump pulse duration. Our experimental observations are in agreement with \textit{ab-initio} calculations and identify the OISTR effect as the underlying mechanism for the emergence of this transient FM state due to the imbalanced population of unoccupied minority states in Mn layers caused by the contribution from the AFM-coupled interfacial Co.


Our sample was grown in an ultra-high vacuum chamber with a base pressure of 1$\times$10$^{-9}$ mbar, on a Cu(001) substrate held at room temperature using e-beam evaporation from a Co rod (99.998\% purity) and Mn flakes (99.99\% purity) in a Ta crucible. We deposited six repetitions of 3 MLs of Co and Mn and on top 14 MLs of Co as a capping layer to prevent the oxidation of the underlying multilayers by residual gas molecules in the ultra-high vacuum. During deposition, the thickness  was determined by the intensity oscillations of diffraction spots in medium-energy electron diffraction while the sample cleanliness was verified by Auger electron spectroscopy. After growth, the sample was stored in a vacuum suitcase with a base pressure better than 2$\times$10$^{-10}$ mbar until its in-vacuum transfer to the magnetic characterization chamber.

 \begin{figure}
 \includegraphics [width=0.45\textwidth]{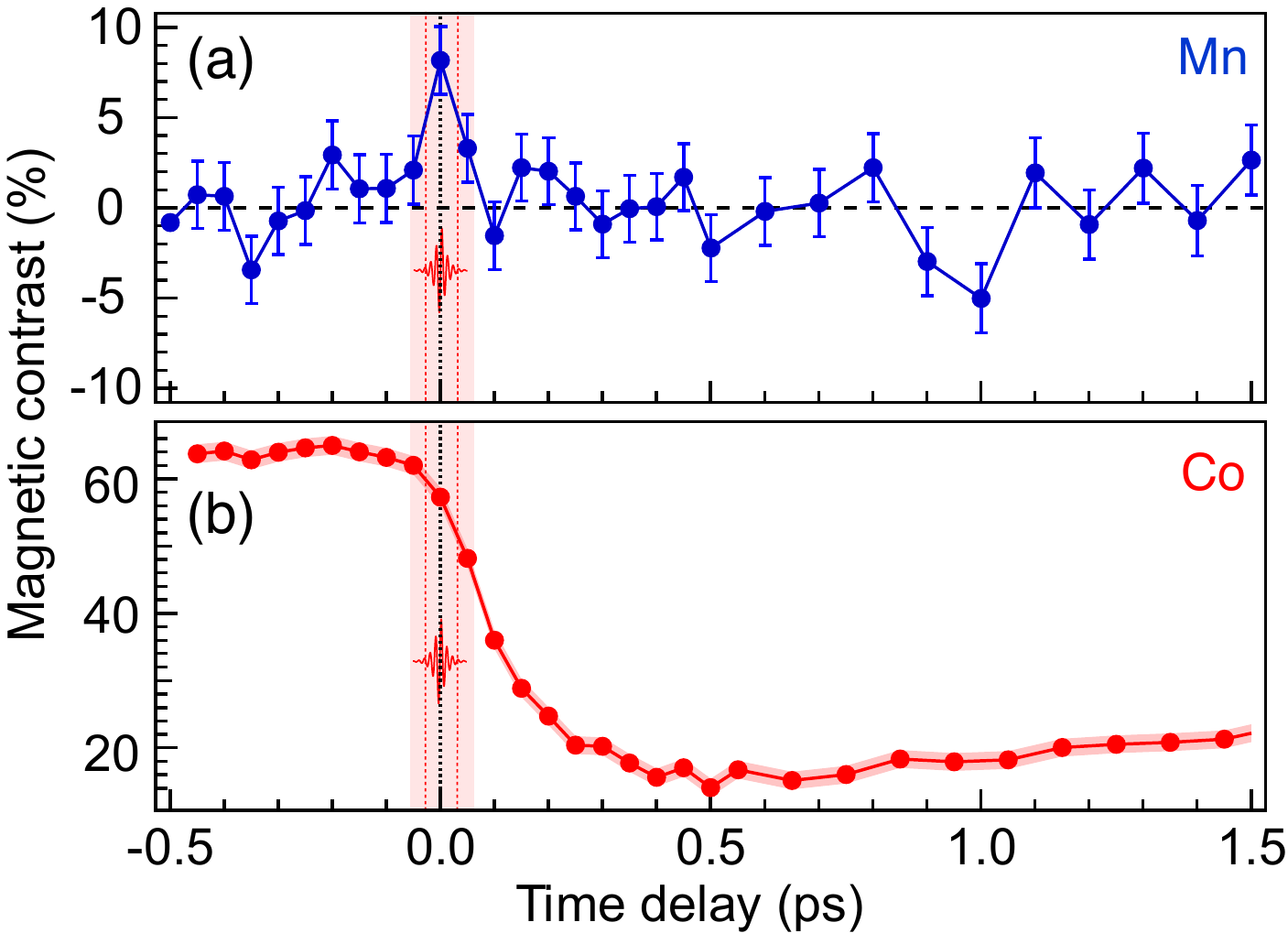}
 \caption{Time-resolved magnetic contrast from the L$_3$ edge of (a) Mn and (b) Co. Blue bars in (a) and red shaded regions in (b) correspond to the statistical errors for the measurements based on Poisson statistics. The zero time delay is defined here at the maximum of the laser pump pulse. The red oscillating and the dashed lines represent the pump pulses and their full width at half maximum, respectively, while the light-red shaded area indicates the experimental time resolution.}
 \label{fig2}
 \end{figure}

We characterized the sample at the FemtoSpeX slicing facility \cite{Holldack:2014fs} at the BESSY II synchrotron of Helmholtz-Zentrum Berlin. Static and dynamic RMXR measurements have been conducted using a magnetic field of 0.2 T with alternating direction between parallel and anti-parallel orientation relative to the x-ray propagation direction and with a fixed x-ray light helicity (see Fig. \ref{fig1}(a)). The time-resolved RMXR measurements have been performed by exciting the sample with linearly-polarized 60-fs laser pulses of 800 nm wavelength and incident fluence F = 12 mJ/cm$^2$, nearly parallel to the x-ray incidence. The magnetic signal was probed with x-ray pulses of 100 fs duration, reaching the sample with a 6 kHz repetition rate, while the pump laser was operated at 3 kHz in order to detect in succession reflected x-rays from the sample with and without laser excitation. The dynamic magnetic signals have been obtained from the difference of the reflected signal with and without laser excitation at the L$_3$ edge of Co and Mn. The total time resolution of our experiment was 120 fs and during all measurements the sample was kept at room temperature. Because of the low intensity of the fs x-ray pulses, our experimental error was determined by photon-counting statistics. Additional characterization of the static magnetic and structural properties of our sample has been performed at the VEKMAG end-station at BESSY II after the dynamic measurements at FemtoSpeX (see supplementary information \cite{SuppInfo}).

\begin{figure*}
\includegraphics [width=0.9\textwidth]{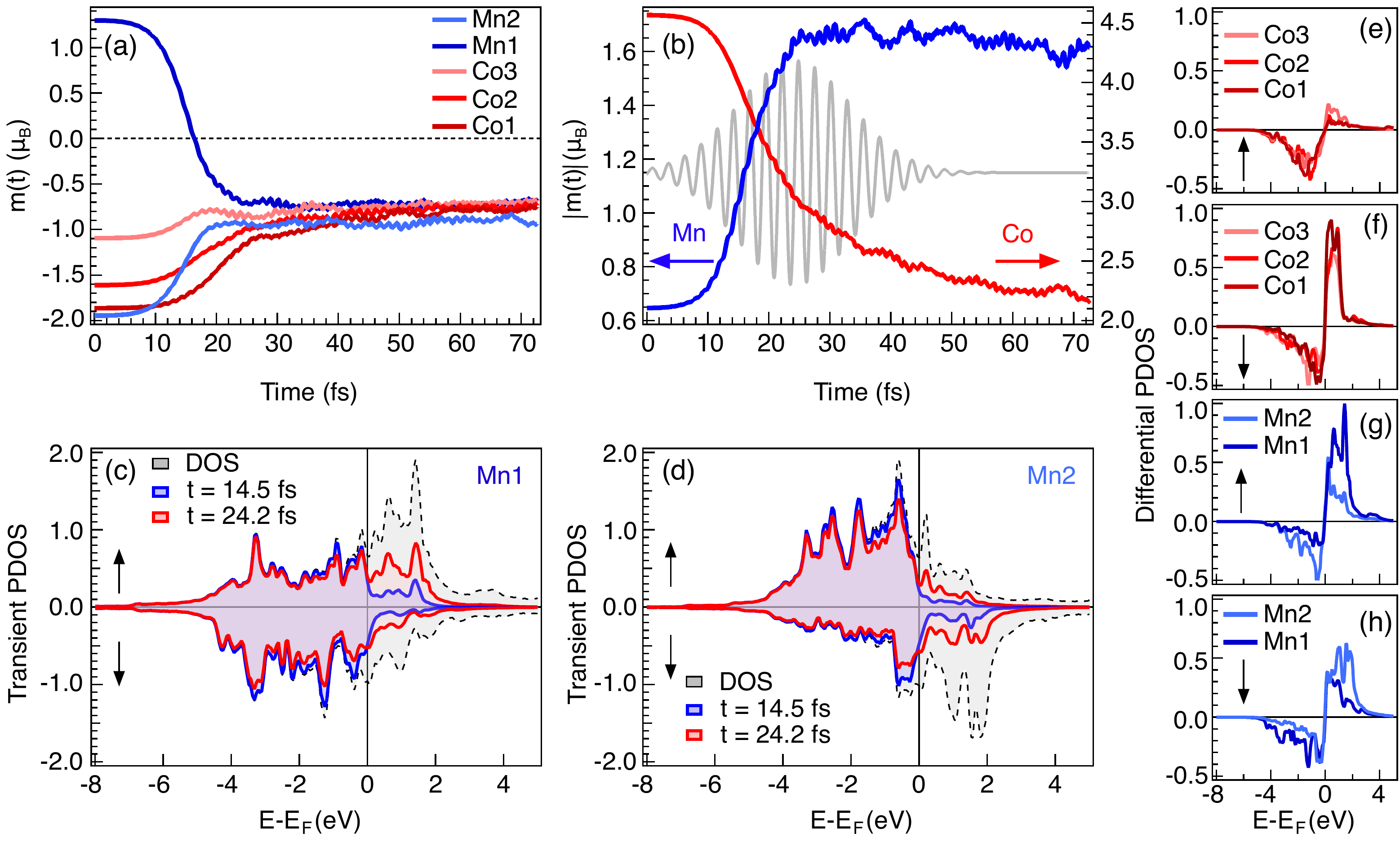}
\caption{TD-DFT calculations for a five-atomic-layer stack comprised of 3 ML Co and 2 ML Mn. (a) Time-dependent magnetic moment of each layer. Note that the stacking order of the layers in the simulation corresponds to the atomic order of the legend and time zero is defined here at the beginning of the excitation pulse, as shown in panel (b). (b) Time-dependent total magnetic moment of all Mn (blue line) and all Co (red line) atoms in the stack. The gray oscillating line corresponds to the temporal profile of the vector potential of the pump pulse. (c), (d) Partial DOS (PDOS) of the Mn atoms at the first (Mn1), second (Mn2) Mn layer, respectively, in states/eV/spin. Black dashed lines represent the full (i.e. occupied and unoccupied) partial DOS of the ground state. Solid blue and red lines represent the transiently occupied/populated part of the PDOS at t = 14.5 fs and t = 24.2 fs, respectively. The zero of the energy scale corresponds to the Fermi energy of the ground state. Up- and down-pointing arrows mark the PDOS for the spin-up (positive values) and spin-down states (negative values), respectively. (e)-(h) Differential partial density of occupied states between t = 24.2 fs and t = 0 fs (equal to the occupied ground state PDOS) for (e) spin up in Co layers, (f) spin down in Co layers, (g) spin up in Mn layers and (h) spin down in Mn layers, respectively.}
\label{fig3}
\end{figure*}


A schematic representation of our sample can be seen in Fig. \ref{fig1}(c). The dominant laminar character of our Co/Mn multilayers has been confirmed by the analysis of x-ray resonant reflectometry oscillations measured at specular geometry. Our sample has a periodicity according to the nominal deposition profile with slight interfacial diffusion. We have calculated the intensity of the pump pulse's electric field as a function of the distance from the sample's surface (see Fig. \ref{fig1}(c)). Our calculations show that 56\% of the incoming infrared light is reflected while $\approx$ 26.6\% is absorbed by the Co layers (13.9\% is the share of the cap layer), 13.3\% by the Mn layers and 4.1\% by the Cu(001) substrate. We estimate that 0.23 and 0.26 photons are absorbed per pulse per Mn and Co atom in our sample, respectively (see supplementary information \cite{SuppInfo}). 

Figure \ref{fig2} displays the time-resolved magnetic signal measured at the L$_3$ resonances of Mn and Co. In static conditions, Co layers do not experience AFM interlayer coupling while the applied magnetic field (0.2 T) is enough to achieve full magnetic saturation. After the excitation pulse, in Fig. \ref{fig2}(b), we detect a strong demagnetization of Co. Fitting the demagnetization curve to an exponential decay function \cite{SuppInfo} results in a demagnetization time constant of 155 $\pm$ 29 fs, in agreement with studies on Co/Pt \cite{Schmising:2015co} and Co/Pd \cite{Vodungbo:2016pd} multilayers.

Most importantly, in Fig \ref{fig2}(a) we observe virtually no magnetic contrast in Mn at negative time delays in accordance with the AFM nature of Mn thin films. Statically, the sample shows a small magnetic dichroism ($\approx$ 1.8\%) antiparallel to the Co magnetization \cite{SuppInfo}, which is below the experimental error in the time-resolved RMXR measurement of Mn. One would expect a higher uncompensated magnetic moment in a perfectly smooth 3-ML Mn film with collinear layer-wise AFM alignment equal to one third of the value the film would have if it was FM aligned. However, the aforementioned condition is relaxed due to imperfections and intermixing at the Mn/Co interface. Moreover, atomic-scale or surface roughness can lead to frustration of the exchange interaction at step edges, leading to canted moments and a deviation from a layer-wise parallel spin structure, resulting in the nearly vanishing static net magnetization we observe in the Mn layers. At the arrival of the excitation pulse, we detect an onset of the magnetic signal of Mn that peaks at 8.2\%. The data points around the peak show a Gaussian trend, consistent with the convolution of our laser-pump and x-ray-probe pulse, with more than 3000 times higher statistical likelihood compared to the average baseline. Right after the pump pulse, the Mn signal returns to its initial ground state value. The maximum lifetime of the transient FM state in Mn is equal or lower than the time resolution of our experiment. 

We attribute the observed transient FM order in Mn to the OISTR effect. We surmise that the FM state in Mn lives roughly as long as the pump pulse is present ($\approx$ 60 fs), given the reported observations of the same effect in Ni/Pt \cite{Siegrist:2019ii} and theoretical considerations \cite{Dewhurst:2018jt}. The estimated lifetime of the transient FM state is consistent with the timescale needed for this excited state to lose coherence due to spin-orbit coupling in an itinerant magnet \cite{Gomez-Abal:2004gl}. We have to stress that the Co magnetic moment sets the preferential orientation of the transient FM magnetization of Mn. We note that, much later, at $\approx$ 1 ps, we see a negative magnetic signal that we can not explain with either OISTR or electronic spin currents. We tentatively assign it to coherent phonons in the multilayer, which change the interatomic magnetic coupling and transiently lead to this signal, for example, by reducing the frustration of Mn magnetic moments at step edges. 

The transient FM alignment in Mn emerges at a delay time when Co has not yet considerably demagnetized \footnote{Magnetization is reduced in Co by $\approx$ 8\% at the peak of the Mn magnetic signal and by $\approx$ 20\% when the Mn magnetic signal is zero}, suggesting the optical nature of the effect. Another mechanism that might play a role in our experiment is superdiffusive transport \cite{Battiato:2012hw, Battiato:2012hw}. However, as the transient FM alignment of Mn layers occurs synchronously with the pump pulse arrival and in the meantime the magnetic signal reduction in Co is small, superdiffusive transport likely does not play a significant role at this early time period.

In order to identify the processes underlying our experimental observations we employed \textit{ab-initio} time-dependent density functional theory (TD-DFT) calculations. We utilized a stack of 2 ML Mn on top of 3 ML Co with an impinging pump pulse with 20 fs full width at half maximum (FWHM) and 19 mJ/cm$^2$ of incident laser pump fluence, as the only input parameters of the calculation. Our model calculations are based on a fully non-collinear version of the Elk code \cite{Dewhurst:2016og,elk}, where electron dynamics after laser excitation is treated by taking into account relativistic effects. Our theoretical approach considers spin and charge currents including superdiffusive currents \cite{Battiato:2010br,Battiato:2012hw}, spin-orbit induced flips, electron-electron scattering and charge- and spin-density waves with unit vectors larger than the size of a unit cell. During these simulations nuclei were kept fixed, as the atomic Hellmann-Feynman forces are very small during the excitation, when our system is in a highly non-equilibrium state, justifying the use of the Born-Oppenheimer approximation. 

We choose to compare our sample with a system with 2 ML Mn on top of 3 ML Co to minimize the total starting magnetization from Mn layers. The simulation of a layered system with zero Mn magnetization as the one studied experimentally would require a large supercell, making the \textit{ab-initio} approach unfeasible. The main conclusions from the calculations do not change, since the parity of Mn layers does not play a role in the emergence of the FM state, as OISTR is mainly an effect between nearest neighbors and decays fast with distance \cite{Dewhurst:2018jt}. Finally, the laser excitation was selected shorter for convenience, as the timescale of the AFM-to-FM transition depends only on the FWHM of the excitation pulse \cite{Dewhurst:2018jt}. As shown in Ref. 25, longer and weaker pulses result in the same physics but with higher computational cost. Therefore, our current approach and conclusions are also valid on the timescale of our experimental excitation.

Our first-principles calculations can qualitatively explain our experimental observations. In Figs. \ref{fig3}(a), (b), we clearly show the transition from AFM to FM alignment of the Mn layers after the arrival of the pump pulse. The onset of the FM state starts right before our pump pulse reaches its half maximum and peaks simultaneously with its vector potential, while Co shows a slower demagnetization in agreement with our experimental observations (see Fig. \ref{fig2}). The underlying mechanism for the transient FM alignment is revealed in Fig. \ref{fig3} (c)-(h), where the unoccupied minority spin density of states (DOS) acts as a sink for excited majority spin electrons from the neighboring Mn layer. The spin swapping between Mn neighbors, facilitated by their AFM coupling, as well as the higher unoccupied state filling of the atoms at the interface (Mn1) from the AFM-coupled reservoir of Co majority electrons drive the transient FM state in Mn \footnote{Integrating the differential partial DOS between t = 0 and 24 fs of Fig. \ref{fig3} (g)-(h) we get +0.87 (-0.29) and +0.60 (-0.20) minority (majority) states for Mn1, Mn2 atoms, respectively. Therefore, the stronger filling of minority states for Mn1 atoms results in the magnetic moment reduction and eventually the reverse of its orientation and the FM alignment with Mn2.}.

In summary, we presented compelling evidence of a transient FM state of AFM Mn in Co/Mn multilayers due to the OISTR effect. The transition is driven by the electric field of the pump pulse in a fs timescale, much faster than the FM-order quenching in Co, while the induced macroscopic magnetic moment of Mn aligns with the adjacent ferromagnet. Our calculations show that the transient FM state originates from the imbalance of intersite transfer of electrons in Mn atoms due to the asymmetry introduced by the Co interface. The lifetime of the effect is comparable to the pump-pulse duration in agreement with theoretical predictions. Our observation validates the hallmark prediction of an important mechanism for ultrafast optical manipulation of magnetic order and most importantly showcases the creation of a transient FM state in a monoelemental antiferromagnet that can play an important role in ultrafast optospintronics.

\begin{acknowledgments}

The authors would like to thank T. Kachel, K. Holldack, R. Mitzner, K. Chen, C. Luo and F. Radu who supported our experiments at the synchrotron facility BESSY II at HZB. We thank HZB for the allocation of synchrotron radiation beamtime. This work was funded by the German Research Foundation (DFG) through CRC/TRR227 projects A03, A04, and A07.

\end{acknowledgments}

%

\end{document}